\documentclass{article}

\usepackage{axodraw}
\usepackage{epsfig}

\newcommand{\NP}[1]{Nucl.\ Phys.\ {\bf #1}}

\begin{document}
\setlength{\unitlength}{1mm}

\begin{titlepage}

\begin{flushright}
LAPTH-845/01\\
LPT-Orsay 01-43\\
May 2001
\end{flushright}
\vspace{1.cm}

\begin{center}
\large\bf
{\LARGE\bf Isolated prompt photon photoproduction at NLO}\\[2cm]
\rm
{M.~Fontannaz$^{a}$, J.~Ph.~Guillet$^{b}$, G.~Heinrich$^{a}$ }\\[.5cm]

{\em $^{a}$Laboratoire de Physique Th\'eorique\footnote{UMR 8627 du
           CNRS} LPT,\\ 
           Universit\'e de Paris XI, B\^atiment 210,\\
           F-91405 Orsay, France} 

\medskip

{\em $^{b}$Laboratoire d'Annecy-Le-Vieux de Physique 
 Th\'eorique\footnote{UMR 5108 du CNRS, associ\'ee \`a 
              l'Universit\'e de Savoie.} LAPTH,}\\
      {\em Chemin de Bellevue, B.P. 110, F-74941 
           Annecy-le-Vieux, France}\\[3.cm]

\end{center}
\normalsize

\begin{abstract}
We present a full next-to-leading order code to calculate the 
photoproduction of prompt photons. The code is a 
general purpose program of "partonic event generator" type
with large flexibility. We study the possibility to constrain 
the photon structure functions and comment on isolation issues.
A comparison to ZEUS data is also shown.

\end{abstract}

\vspace{3cm}

\end{titlepage}

\section{Introduction}

High energy electron-proton scattering at the DESY $ep$ collider HERA 
is dominated by photoproduction processes, where the electron 
is scattered at small angles, emitting a quasireal photon which 
scatters with the proton. These processes are of special interest 
since they are sensitive to both the partonic structure of the photon 
as well as of the proton. 
In particular, they offer the possibility  to constrain the 
(presently poorly known) gluon distributions in the photon, 
since in a certain kinematical region the subprocess 
$q g\to \gamma q$, where the gluon is stemming from a resolved photon,
is dominating. Up to now, the experimental errors were too large to 
discriminate clearly between different sets of gluon distributions in 
the photon, but a high statistics analysis of the 1996-2000 HERA data on   
prompt photon photoproduction   
announced by the ZEUS collaboration will shed new light on this issue. 


The calculation of higher order corrections to the Compton process 
$\gamma q\to \gamma q$
has been initiated some time ago
~\cite{Duke:1982bj}--\cite{Gordon:1995km}. 
The  most recent calculations for prompt photon photoproduction 
have been done by Gordon/Vogelsang\cite{Gordon:1995km} for isolated 
prompt photon production, Gordon~\cite{Gordon} for photon plus jet 
production and by the group
Krawczyk/Zembrzuski~\cite{Krawczyk} for both the inclusive case and  
$\gamma$+jet.  However, all of these calculations contain
certain drawbacks. In~\cite{Gordon:1995km}, isolation is implemented 
by adding a subtraction term evaluated in the collinear approximation 
to the fully inclusive cross section. 
The programs of \cite{Gordon} and \cite{Krawczyk} do not contain  
the full set of NLO corrections. In \cite{Gordon}, those parts where 
the final state photon comes from fragmentation of a hard parton 
were included only at leading order, arguing that isolation cuts
will suppress the fragmentation component in any case to 
a large extent. Moreover, the box contribution has not been included. 
In \cite{Krawczyk}, higher order corrections are included only for 
the case where initial and final state 
photons are both direct. So not only the contributions from fragmentation, 
but also the case where the initial photon is resolved 
are included at Born level only. 
However, the box contribution has been taken into account.

The calculation presented in this paper takes into account 
the full NLO corrections to all four subparts. 
The corresponding matrix elements already have been calculated and tested in 
previous works~\cite{Aurenche:1984hc,ellissexton,Aurenche:1987ff}. 
A major advantage of the present code is also given by the fact that it 
is constructed as a "partonic event generator" and as such is very flexible.
Various sorts of observables matching a particular experimental
analysis can be defined and histogrammed for an event sample 
generated once and for all. 
This strategy already has been applied to construct NLO codes for 
$\gamma\gamma$ production (DIPHOX)~\cite{Binoth:2000qq} and 
one or two jets photoproduction~\cite{Aurenche:2000nc}.

The paper is organized as follows. In section 2 we first  
describe the theoretical framework and  the treatment of the 
infrared singularities. Then we discuss the implementation of isolation 
cuts and outline the structure of the code. 
Section 3 is devoted to phenomenology. 
We study the effect of isolation, determine the kinematic region 
which is most sensitive to the gluon distribution in the photon and
illustrate the sensitivity of the cross section to the energy of the 
incoming photon.  
We give results for inclusive isolated prompt photon production 
and compare with a recent analysis of ZEUS data~\cite{Breitweg:2000su},
 before we come to the conclusions in section 4.

\section{Theoretical formalism and description of the method}

In this section the general framework for prompt photon 
photoproduction will be outlined. 
We will review the contributing subprocesses, the treatment of 
infrared singularities and the implementation of isolation cuts. 

\subsection{The subprocesses contributing at NLO}

The inclusive cross section for $e p\to\gamma X$ can symbolically 
be written as a convolution of the parton densities of the incident particles
(resp. fragmentation function for an outgoing parton  fragmenting into a
photon) with the partonic cross section $\hat \sigma$
  
\begin{eqnarray}
d\sigma^{ep\to\gamma X}(P_p,P_e,P_{\gamma})&=&\sum_{a,b,c}\int dx_e\int d x_p\int
dz\, F_{a/e}(x_e,M)F_{b/p}(x_p,M_p)\nonumber\\
&&d\hat\sigma^{ab\to c
X}(x_pP_p,x_eP_e,P_{\gamma}/z,\mu,M,M_p,M_F)\,D_{\gamma/c}(z,M_F)
\label{dsigma}
\end{eqnarray}
where $M,M_p$ are the initial state factorization scales, $M_F$ the 
final state factorization scale  and $\mu$ the
renormalization scale. 

The subprocesses contributing 
to the partonic reaction $ab \to c X$ can be divided into four 
categories which will be denoted by
1.~direct direct \, 2.~direct fragmentation 
\, 3.~resolved direct \, 4.~resolved fragmentation.  
The cases "direct direct" and "resolved direct" correspond to 
$c=\gamma$ and $D_{\gamma/c}(z,M_F)=\delta_{c\gamma}\delta(1-z)$ 
in (\ref{dsigma}), that is, the prompt\footnote{By "prompt" we
mean that the photon is not produced from the decay of light mesons.} 
photon is produced directly in the hard subprocess. 
The cases with "direct" attributed to the initial state photon 
correspond to  $a=\gamma$, with $F_{\gamma/e}$ 
approximated by the Weizs\"acker-Williams formula for the spectrum of 
the quasireal photons 
\begin{equation}
f^e_{\gamma}(y) = \frac{\alpha_{em}}{2\pi}\left\{\frac{1+(1-y)^2}{y}\,
\ln{\frac{Q^2_{\rm max}(1-y)}{m_e^2y^2}}-\frac{2(1-y)}{y}\right\}\;.
\label{ww}
\end{equation}

The "resolved" contributions are characterized by a resolved photon in the
initial state where a parton stemming from the  photon 
instead of the photon itself participates in the hard subprocess. 
In these cases, $F_{a/e}(x_e,M)$ is given by a convolution of the 
Weizs\"acker-Williams spectrum with the parton distributions 
in the photon:

\begin{equation}
F_{a/e}(x_e,M)=\int_0^1 dy \,dx_{\gamma}\,f^e_{\gamma}(y) \,
F_{a/\gamma}(x_{\gamma},M)\,\delta(x_{\gamma}y-x_e)
\end{equation}

\begin{figure}
\begin{center}
\begin{picture}(100,80)(0,-30)
\ArrowLine(-30,20)(0,40)
\ArrowLine(-60,20)(-30,20)
\Photon(0,0)(-30,20){2}{5}
\Line(0,0)(30,20)
\Line(0,0)(0,-40)
\Line(0,-40)(-30,-60)
\Photon(0,-40)(30,-60){2}{5}
\BCirc(-32,-62){7}
\Line(-60,-60)(-39,-60)
\Line(-60,-64)(-39,-64)
\Line(-25,-62)(-8,-66)
\Line(-25,-64)(-10,-70)
\Line(-25,-66)(-12,-74)
\Text(-20,-18)[t]{p}
\Text(-20,10)[t]{$e$}
\Text(-4,7)[t]{$\gamma$}
\ArrowLine(170,20)(200,40)
\ArrowLine(140,20)(170,20)
\Photon(200,0)(170,20){2}{5}
\Line(200,0)(230,20)
\Line(200,0)(200,-40)
\Line(200,-40)(170,-60)
\Gluon(200,-40)(230,-60){2}{5}
\BCirc(234,-63){4}
\Line(238,-60)(260,-62)
\Line(238,-62)(260,-67)
\Photon(238,-65)(260,-79){2}{3}
\BCirc(168,-62){7}
\Line(140,-60)(161,-60)
\Line(140,-64)(161,-64)
\Line(175,-62)(192,-66)
\Line(175,-64)(190,-70)
\Line(175,-66)(188,-74)
%
\end{picture}
\end{center}
\caption{Examples of direct direct and direct fragmentation contributions at leading order}
\label{fig1}
\end{figure}
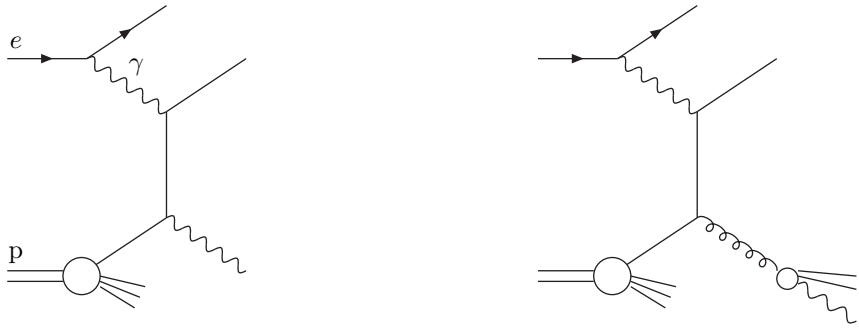

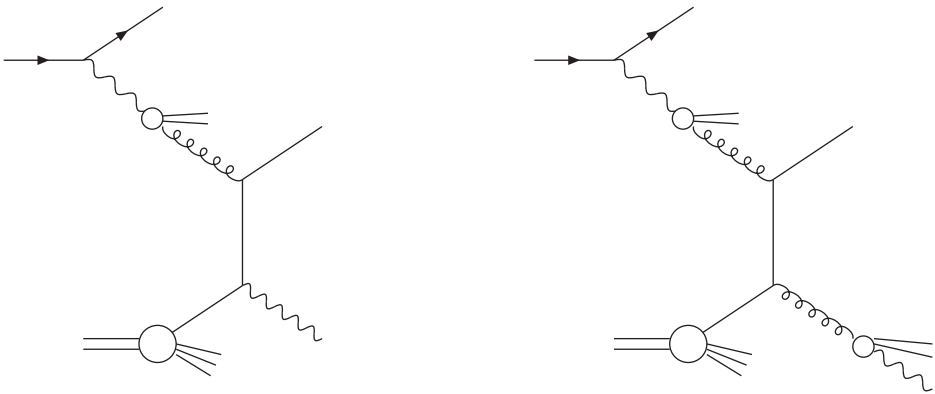
\begin{figure}
\begin{center}
\begin{picture}(100,80)(0,-30)
\ArrowLine(-60,45)(-30,65)
\ArrowLine(-90,45)(-60,45)
\Photon(-60,45)(-36,26){2}{3}
\BCirc(-34,23){4}
\Line(-30,22)(-13,21)
\Line(-30,24)(-13,25)
\Gluon(0,0)(-30,20){2}{5}
\Line(0,0)(30,20)
\Line(0,0)(0,-40)
\Line(0,-40)(-30,-60)
\Photon(0,-40)(30,-60){2}{5}
\BCirc(-32,-62){7}
\Line(-60,-60)(-39,-60)
\Line(-60,-64)(-39,-64)
\Line(-25,-62)(-8,-66)
\Line(-25,-64)(-10,-70)
\Line(-25,-66)(-12,-74)
%
\ArrowLine(140,45)(170,65)
\ArrowLine(110,45)(140,45)
\Photon(140,45)(164,26){2}{3}
\BCirc(166,23){4}
\Line(170,22)(187,21)
\Line(170,24)(187,25)
\Gluon(200,0)(170,20){2}{5}
\Line(200,0)(230,20)
\Line(200,0)(200,-40)
\Line(200,-40)(170,-60)
\Gluon(200,-40)(230,-60){2}{5}
\BCirc(234,-63){4}
\Line(238,-60)(260,-62)
\Line(238,-62)(260,-67)
\Photon(238,-65)(260,-79){2}{3}
\BCirc(168,-62){7}
\Line(140,-60)(161,-60)
\Line(140,-64)(161,-64)
\Line(175,-62)(192,-66)
\Line(175,-64)(190,-70)
\Line(175,-66)(188,-74)
%
\end{picture}
\end{center}
\caption{Examples of resolved direct and resolved fragmentation contributions at leading order}
\label{fig2}
\end{figure}

Examples of diagrams contributing at Born level to the four categories 
above are shown in Figs.~\ref{fig1} and \ref{fig2}. 

\begin{figure}
\begin{center}
\begin{picture}(70,50)(-30,-10)
\Line(0,0)(0,50)
\Line(0,50)(50,50)
\Line(50,50)(50,0)
\Line(0,0)(50,0)
\Photon(-40,50)(0,50){2}{5}
\Gluon(-40,0)(00,00){2}{5}
\Gluon(50,50)(90,50){2}{5}
\Photon(50,0)(90,0){2}{5}
\end{picture}
\end{center}
\caption{The box contribution}
\label{box}
\end{figure}
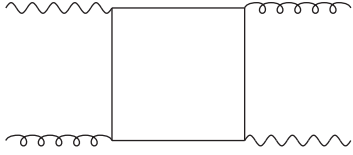

In the case of the  "direct direct" part, only the 
Compton process $\gamma q \to \gamma q$ contributes at leading order, at NLO 
the ${\cal O}(\alpha_s)$ corrections from $\gamma q \to \gamma q g$\quad 
resp. $\gamma g \to \gamma q \bar q$ and the corresponding virtual corrections 
contribute.
We also included the box 
contribution (Fig.~\ref{box}) into the "direct direct" part since it is known 
to be sizeable~\cite{Aurenche:1992sb}, although it is formally a NNLO 
contribution.

In the "direct fragmentation" part, the final state photon comes from the 
fragmentation of a hard parton participating in the short distance subprocess.
From a technical point
of view, a final state quark-photon collinear singularity appears in the
calculation of the  subprocess $\gamma g  \rightarrow 
\gamma q\bar q$.
At higher orders, final state multiple collinear singularities
appear in any subprocess where a high $p_{T}$ parton (quark or gluon)
undergoes a cascade of successive collinear splittings ending up with a
quark-photon splitting. These singularities are factorized to all orders in 
$\alpha_s$  and absorbed, at some arbitrary fragmentation scale $M_F$, 
into quark and gluon fragmentation functions to a photon,  
$D_{\gamma/c}(z,M_F^{2})$. 
When the fragmentation
scale $M_F$, chosen of the order of the hard scale of the subprocess, is
large compared to any typical hadronic scale $\sim 1 $~GeV, these functions
behave roughly as $\alpha/\alpha_s(M_F^{2})$. Then a power counting argument
tells that these fragmentation contributions are asymptotically of the same order in
$\alpha_s$ as the Born term. A consistent NLO calculation thus requires 
the inclusion of the ${\cal O}(\alpha_s)$ corrections to these contributions.
 
Note that the singularity appearing in the process 
$\gamma g  \rightarrow \gamma q\bar q$ when the final state photon 
is emitted by the quark and becomes collinear, 
is subtracted and absorbed by the 
fragmentation function at the scale $M_F$, as explained above. 
Therefore both the "direct direct" and the "direct fragmentation" parts
{\it separately}  
depend strongly on $M_F$ and the attribution of the finite terms to 
either of these parts is scheme dependent. 
Only in the sum of these parts the $M_F$ dependence 
flattens as expected.  

The collinear singularities appearing at NLO if the {\it incident} photon 
splits into a collinear $q \bar q$ pair are absorbed into the functions 
$F_{q/\gamma}(x_{\gamma},M)$ at the factorization scale $M$. 
(Analogous for the proton distribution functions $F_{b/p}(x_p,M_p)$\,; 
we will set $M_p=M$ in the following.) 
Thus, by the same reasoning 
as above for the final state, the "initial direct" and "initial resolved"
parts separately show a strong dependence on $M$ which cancels out 
in the sum.
Therefore it has to be stressed that only the sum over all four parts 
has a physical meaning. 
Figure~\ref{varyM} illustrates these cancellation mechanisms. 

\begin{figure}[htb]
\begin{center}
\mbox{\epsfig{file=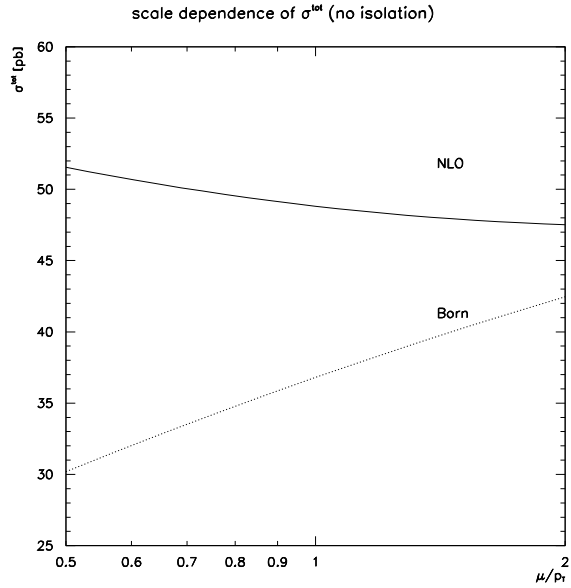,height=8.3cm}}
\end{center}
\caption{Dependence of the total cross section on scale variations. $\mu=M=M_F$
is varied between  $\mu=p_T^{\gamma}/2$ and 
$\mu=2\,p_T^{\gamma}$.}
\label{varyscales}
\end{figure}

\begin{figure}[htb]
\begin{center}
\mbox{\epsfig{file=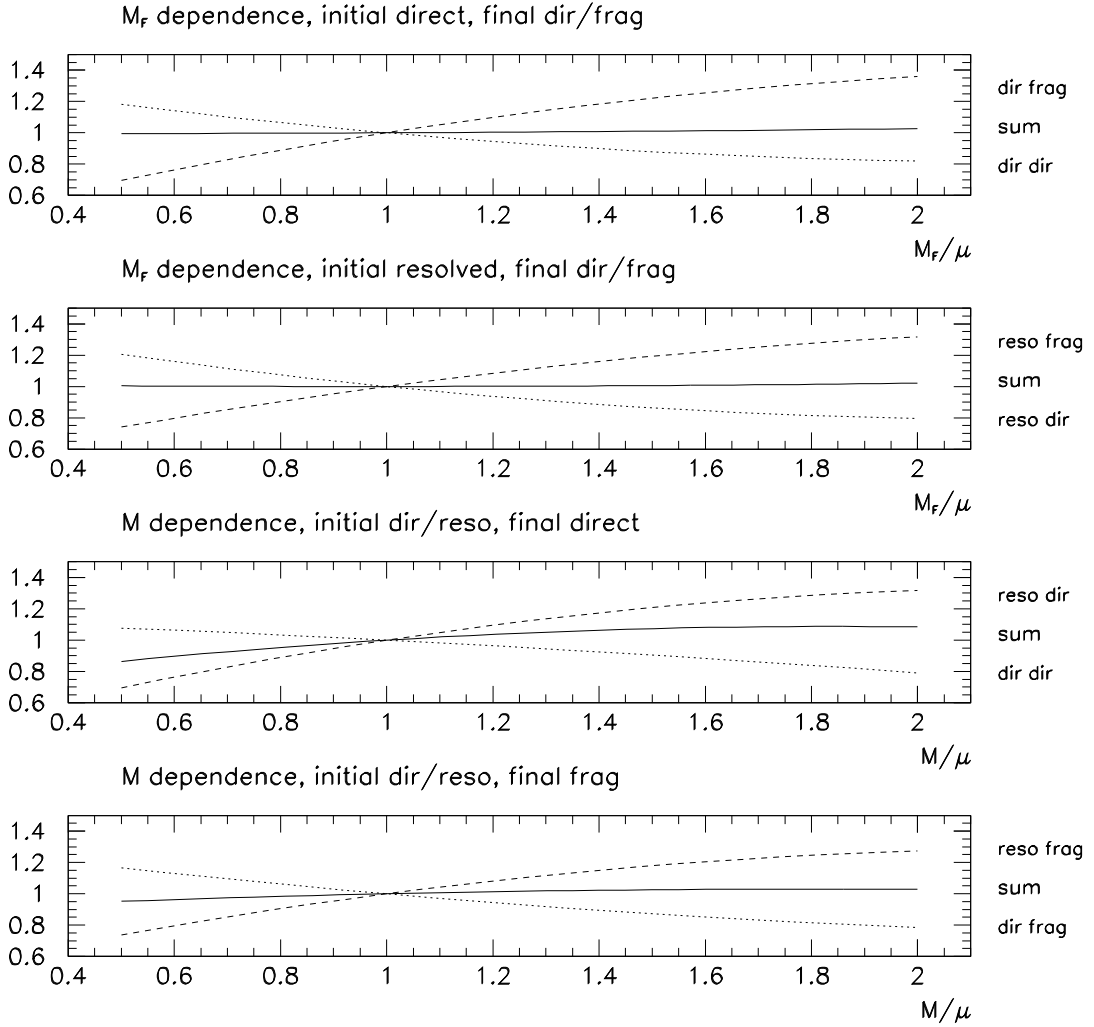,height=15cm}}
\end{center}
\caption{Cancellation of the leading dependence on the fragmentation
scale $M_F$ between contributions from direct and fragmentation final states, 
and on the factorization scale $M$ between parts with 
direct and resolved initial state.  The results are normalized to the 
total cross section at $M_F=M=\mu=p_T^{\gamma}$.}
\label{varyM}
\end{figure}

The overall reduction of the scale dependence when going from leading 
to next-to-leading order can be seen in Fig.~\ref{varyscales}. 
The scales $M_F$ and $M$
have been set equal to $\mu$,  and $\mu$ has been varied between 
$\mu=p_T^{\gamma}/2$ and $\mu=2\,p_T^{\gamma}$. 
One can see that the NLO cross section is much more stable 
against scale variations, it varies by less than 10\% in this $\mu$ range. 

\clearpage

\clearpage

\subsection{Treatment of infrared singularities}\label{ir}

There are basically two methods to isolate the infrared singularities 
appearing in the calculation at NLO: The phase space slicing
method~\cite{slice} and the subtraction method~\cite{subt}. 
The method used here follows the approach of 
\cite{Binoth:2000qq,Chiappetta:1996wp} 
which combines these two techniques. We will outline the strategy 
only shortly, for more details we refer to \cite{Binoth:2000qq}. 

For a generic reaction $1 + 2 \rightarrow 3 + 4 + 5$, at least two particles of the
final state, say 3 and 4, have a high $p_{T}$ and are well separated in phase
space, while the last one, say 5, can be soft, and/or collinear to either of the
four others. 
In order to extract these singularities, the phase space is cut into two
regions:
\begin{itemize}
\item[--] part I where 
the norm $p_{T 5}$ of the transverse momentum of particle 5 is required to be
less than some arbitrary value $p_{Tm}$ taken to be small compared to the other
transverse momenta. This cylinder contains the infrared  and the initial state
collinear singularities. It also contains a small fraction of the final state
collinear singularities. \\
\item[--] parts II a(b) where 
the transverse momentum vector of the particle 5 is required to have a norm
larger than $p_{Tm}$, and to belong to a cone $C_3 (C_4)$ about the direction of 
particle $3 (4)$, defined by  
$(\eta_5-\eta_i)^2+(\phi_5-\phi_i)^2~\leq~R_{th}^2\; (i=3,4)$, with $R_{th}$ some small 
arbitrary number. $C_3 (C_4)$ contains the final state collinear singularities 
appearing when $5$ is collinear to $3 (4)$.\\
\item[--] part II c where $p_{T 5}$ is required to have a
norm larger than $p_{Tm}$, and to belong to neither of the two cones $C_3$, 
$C_4$. This slice yields no divergence, and can thus be treated directly in
$4$ dimensions.
\end{itemize}
The contributions from regions I and IIa,b are calculated analytically in 
$d=4-2\epsilon$ dimensions and then combined with the corresponding virtual 
corrections such that the infrared singularities cancel, except for the initial
(resp. final) state collinear singularities, which are factorized and absorbed 
into the parton distribution (resp. fragmentation) functions. 

\begin{figure}[htb]
\begin{center}
\mbox{\epsfig{file=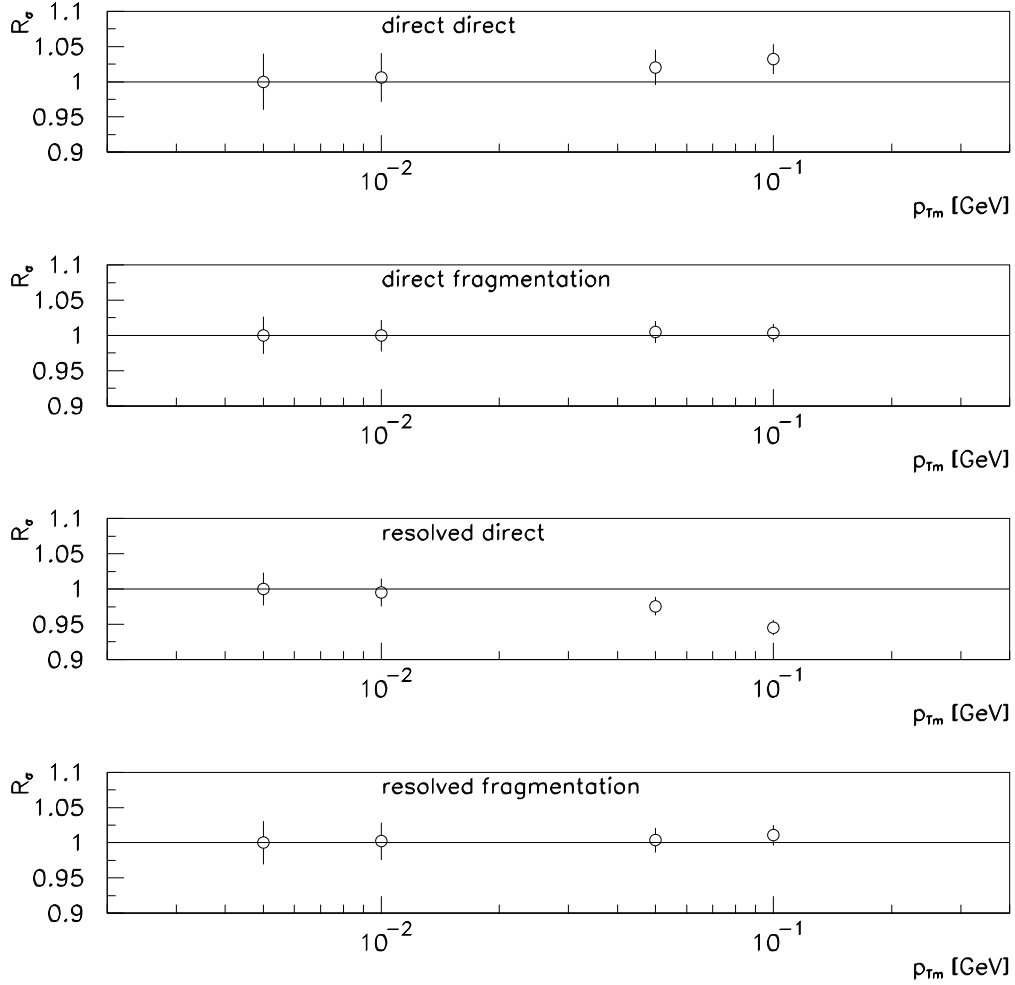,height=15cm}}
\end{center}
\caption{Dependence of the total cross section on variations of the slicing
parameter $p_{Tm}$. $R_{\sigma}$
denotes the total (nonisolated) cross section normalized to the total 
cross section evaluated with $p_{Tm}=0.005$, 
$R_{\sigma}(p_{Tm})=\sigma^{tot}(p_{Tm})/\sigma^{tot}(p_{Tm}=0.005)$.
One can see that there is a plateau where the cross 
section is fairly insensitive to variations of $p_{Tm}$. The same study 
has been made for the dependence on $R_{th}$, but there the cross section 
is completely stable within the numerical errors since the $R_{th}$ 
dependence has been kept exactly in all parts of the matrix element. }
\label{varyptm}
\end{figure}

After the cancellation, the finite remainders of the soft and collinear 
contributions in parts I and II\,a,b,c separately depend on large logarithms 
$\ln p_{Tm}, \ln^2 p_{Tm}$  and $\ln R_{th}$. 
When combining the different parts, the following cancellations of
the $p_{Tm}$ and $R_{th}$ dependences occur:\\
 In part I, the finite terms are approximated by 
collecting all the terms depending logarithmically on $p_{Tm}$ and 
neglecting the terms proportional to powers of $p_{Tm}$.  
 On the contrary, the $R_{th}$ dependence in the conical parts II\,a and II\,b, is kept exactly. This means that an exact cancellation of the 
dependence on the unphysical parameter $R_{th}$  between 
part II\,c and parts II\,a,b  occurs, 
whereas the cancellation of the unphysical parameter
$p_{Tm}$  between parts II\,c, II\,a,b  and part I is only approximate. The
parameter $p_{Tm}$ must be chosen small enough with respect to $p_{T}^{\gamma}$ 
in order that the neglected terms can be safely dropped out. 
On the other hand, it cannot be chosen too small since otherwise numerical 
instabilities occur.  
We have investigated the stability of the cross section by varying  $p_{Tm}$ and 
$R_{th}$ between $0.005$ and  $0.1$ (see Figure \ref{varyptm}) and accordingly 
chosen the optimal values $p_{Tm}=0.05$\,GeV, $R_{th}=0.05$.

\clearpage

\subsection{Implementation of isolation cuts}\label{iso}

In order to single out the prompt photon events from the huge background
of secondary photons produced by the decays of $\pi^0,\eta,\omega$ mesons,
isolation cuts have to be imposed on the photon signals in the experiment. 
A commonly used isolation criterion is the following\footnote{
A more sophisticated criterion  has been 
proposed in \cite{frixione}, in which the veto on accompanying
hadronic transverse energy is the more severe, the closer the corresponding
hadron to the photon direction. It has been designed to make the
``fragmentation" contribution vanish in an infrared safe way.}:
A photon is isolated if, inside a cone centered around the photon direction 
in the rapidity and azimuthal angle plane, the amount of hadronic transverse 
energy
$E_T^{had}$ deposited is smaller than some value $E_{T\, max}$ fixed by the
experiment:
\begin{equation}\label{criterion}
\left.
\begin{array}{rcc} 
\left(  \eta - \eta_{\gamma} \right)^{2} +  \left(  \phi - \phi_{\gamma} \right)^{2}  
& \leq  & R_{\mathrm{exp}}^{2} \\
E_T^{had} & \leq & E_{T\, max}
\end{array}
\right\} 
\end{equation}
Following the conventions of the ZEUS collaboration, we used 
$E_{T\, max}=\epsilon \,p_T^{\gamma}$ with 
$\epsilon=0.1$ and $R_{\mathrm{exp}}$ = 1.
Isolation not only reduces the background from secondary photons, but also 
substantially  reduces the fragmentation components, as will be illustrated 
in section \ref{incl}. 

Furthermore, it is important to note that the isolation parameters must be
carefully fixed in order to allow a comparison between data and perturbative 
QCD calculations. Indeed a part of the hadronic energy measured in the cone 
may come from the underlying event; therefore even the direct contribution 
can be cut by the isolation condition if the latter is too stringent. 
Let us estimate the importance of this effect and assume that the underlying
event one-particle inclusive distribution is given by
\begin{equation}
\frac{dn^{(1)}}{p_T\,dp_T\,d\eta\, d\phi}=\frac{\bar{n}}{2\pi}\frac{4}{\langle
p_T\rangle^2}\,e^{-\frac{2 p_T}{\langle
p_T\rangle}}\;,
\end{equation}
$n^{(1)}$ being normalized to $\bar{n}$ particles per unit of rapidity. 
The probability that the isolation condition is fulfilled by a 
particle from an underlying event is 
\begin{eqnarray}
n_{isol}^{(1)}&=&\int_{cone}d\phi\, d\eta\int_{E_{T\, max}}^{\infty}p_T dp_T\,
\frac{dn^{(1)}}{p_T\,dp_T\,d\eta \,d\phi}\nonumber\\
&=&R_{\mathrm{exp}}^{2}\frac{\bar{n}}{2}\left(1+\frac{2 E_{T\, max}}{\langle
p_T\rangle}\right)\,e^{-\frac{2 E_{T\, max}}{\langle p_T\rangle}}
\label{estim}
\end{eqnarray}
With the ZEUS isolation parameters, $E_{T\, max}=0.5$\,GeV for a photon of 
$p_T^{\gamma}=5$\,GeV. Using $\bar{n}=3$ and $\langle p_T\rangle \approx 0.35$\,GeV 
extracted from \cite{capella}, one obtains 
$$n_{isol}^{(1)}\approx 0.33 $$
This estimation is very rough and underestimates the true effect 
because there is also a non-negligible probability to fulfill the 
isolation condition with two underlying particles falling into the cone. 
Only a detailed Monte Carlo description of the underlying events can allow 
a reliable estimate of this non-perturbative effect. Here we just note that the
cut put by ZEUS ($E_{T\, max}\approx 0.5$\,GeV) is likely to be too low 
to eliminate any underlying event contamination and therefore makes a
comparison between the partonic level QCD predictions and the (hadronic level) 
data difficult.


\subsection{Features of the code}

The code consists of four subparts corresponding to each of the four 
categories of subprocesses. For each category, the functions corresponding to 
the parts I, II\,a,b,c described in section \ref{ir} are integrated 
separately with the numerical 
integration package BASES\cite{bases}. Based on the grid produced by this
integration, partonic events are generated with SPRING\cite{bases} and 
stored into an NTUPLE or histogrammed directly. 
It has to be emphasized that we generate
final state {\it partonic} configurations. 
Hence this type of program does not provide an
exclusive portrait of final states as given by hadronic event generators
like PYTHIA  or HERWIG. On the other hand, the
latter are only of some improved leading logarithmic accuracy. 
The information stored in the NTUPLE are the 4-momenta of the outgoing 
particles, their types (i.e. quark, gluon or photon), 
the energy of the incident photon and, 
in the fragmentation cases, the longitudinal
fragmentation variable associated with the photon from fragmentation. 
Furthermore a label is stored that allows to identify the 
origin of the event, e.g. if it  comes from a $2\to 2$ or a $2\to 3$ 
process. 
Based on the information contained in these NTUPLES, suitable observables can
be defined and different jet algorithms can be studied. 
The isolation cuts are included already at the integration 
level, but the user of the program can  turn isolation 
on or off and vary the input parameters for the isolation cut at will.

\section{Numerical results and comparison to ZEUS data}

In this section we present some numerical results for isolated
prompt photon production. We restrict ourselves to the inclusive 
case, photon + jet production will be discussed in detail 
in a forthcoming publication. 

For the parton distributions in the
proton we take the MRST2~\cite{Martin:2000ww} parametrization. 
Our default choice for the photon distribution functions is 
AFG~\cite{Aurenche:1994in}, 
for  comparisons we also used the GRV~\cite{Gluck:1992ee} 
distributions transformed to the $\overline{\rm{MS}}$ scheme. 
For the fragmentation functions we use
the parametrization of Bourhis et al~\cite{Bourhis:1998yu}. 
We take $n_f=4$ flavors, the contributions from the $b$ quark
can be neglected since the typical energy scale of the partonic process 
is too close to  the $b$-production threshold.
For $\alpha_s(\mu)$ we use an exact 
solution of the two-loop renormalization group
equation, and not an expansion in log$(\mu/\Lambda)$. 
Unless stated otherwise, the scale choices $M=M_F=\mu=p_T^{\gamma}$ 
have been used. 
The rapidities refer to the $e\,p$ laboratory frame, with the HERA
convention that the proton is moving towards positive rapidity. HERA operates
with an electron energy of 27.5 GeV and a proton energy of 820 GeV.

Following the ZEUS collaboration~\cite{Breitweg:2000su}, 
the parameters for the Weizs\"acker-Williams 
spectrum are $Q^2_{\rm{max}}$=1\,GeV and the photon energy $y=E_{\gamma}/E_e$
has been restricted to the range $0.2<y<0.9$. 
Note that experimentally, 
the energy of the incoming photon in photoproduction 
processes is reconstructed from the final hadron 
energies with the Jacquet-Blondel method, 
\begin{equation}
y_{JB}=\frac{\sum(E-p_z)}{2E_e}
\label{yjb}
\end{equation}
where the sum is over all calorimeter cells, $E$ is the energy 
deposited in the cell and $p_z=E\cos\theta$. In order to obtain the "true"
photon energy $y$, corrections for detector effects and 
energy calibration have to be applied to $y_{JB}$.
These corrections are assumed to be uniform over the whole $y$ range and 
enter into the experimental systematic error. However, as the 
background varies with the photon energy $y$, these corrections may 
not be uniform. 
It has to be emphasized that the cross section is very sensitive to a 
variation of the energy range of the photon. (See Figure \ref{varyeta} 
and discussion below.)

\subsection{Numerical results for inclusive prompt photon production}\label{incl}

If not stated otherwise, all plots showing the photon rapidity ($\eta^{\gamma}$)
dependence are integrated over 5\,GeV $<p_T^{\gamma}<10$\,GeV and
0.2 $<y=E_{\gamma}/E_e<$ 0.9. 

\begin{figure}[htb]
\begin{center}
\mbox{\epsfig{file=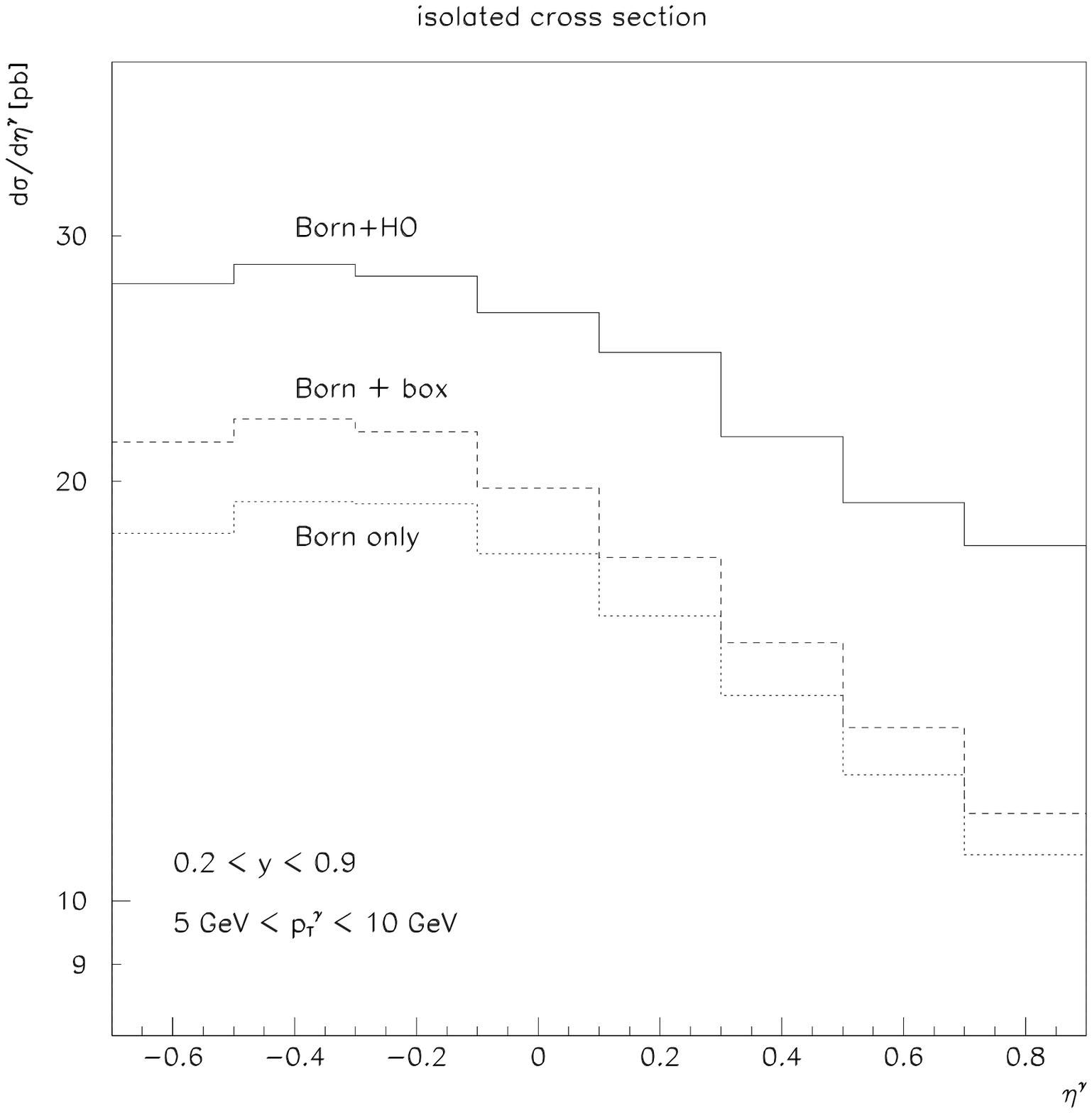,height=15cm}}
\end{center}
\caption{Comparison of NLO to LO result for the photon rapidity distribution.}
\label{bornhoall}
\end{figure}

Figure~\ref{bornhoall} shows a comparison of the NLO 
to the leading order result for the isolated cross section 
$d\sigma/d\eta^{\gamma}$
The importance of the box contribution is clearly visible. 
The higher order corrections enhance the isolated cross section by about 40\%. 

\begin{figure}[htb]
\begin{center}
\mbox{\epsfig{file=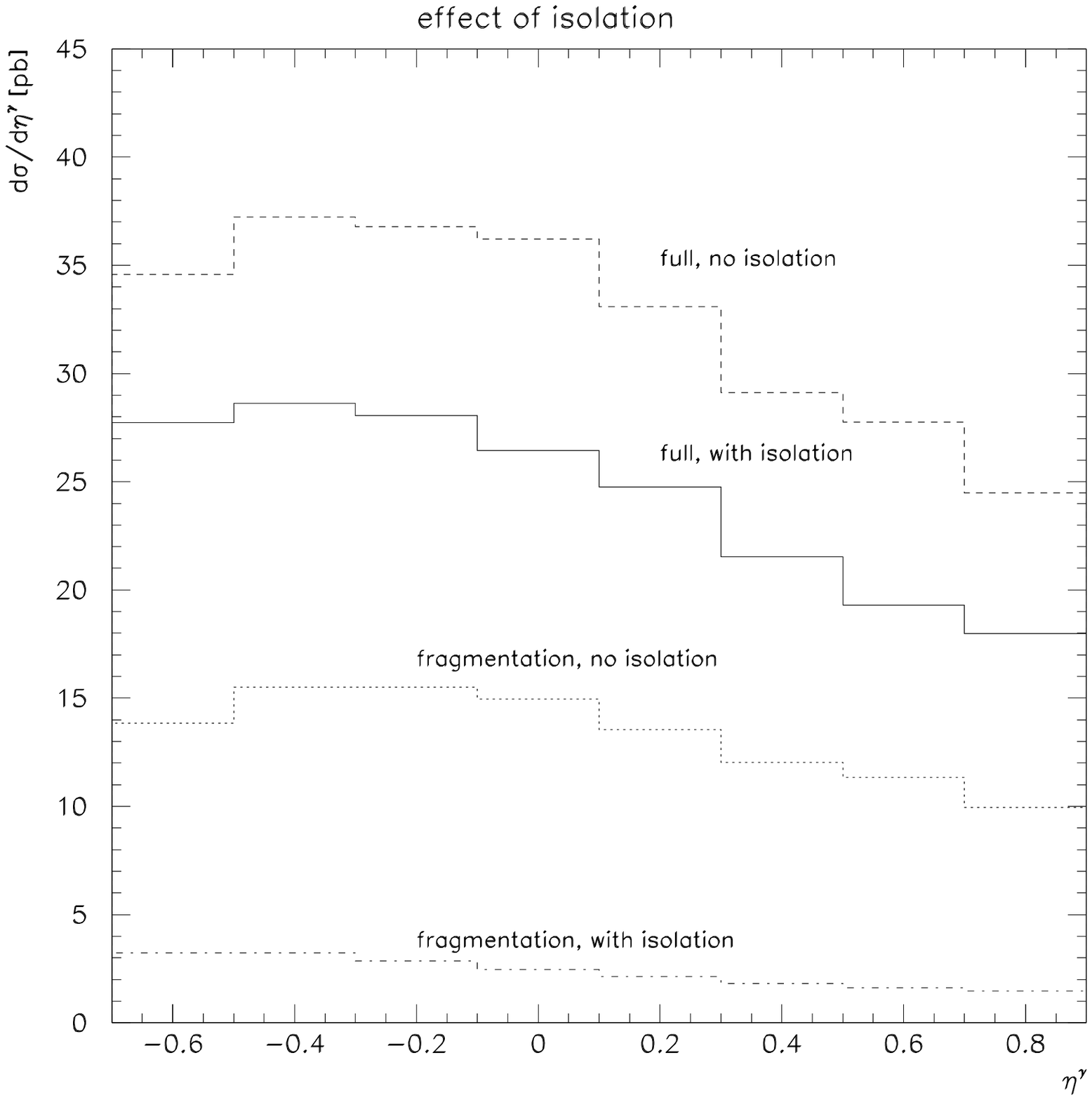,height=15cm}}
\end{center}
\caption{Effect of isolation on the photon rapidity distribution 
$d\sigma^{ep\to\gamma X}/d\eta^{\gamma}$ for the 
full cross section and for the fragmentation components separately. 
Isolation with $\epsilon=0.1,
R_{\mathrm{exp}}=1$.}
\label{isofull/frag}
\end{figure}

\begin{figure}[htb]
\begin{center}
\mbox{\epsfig{file=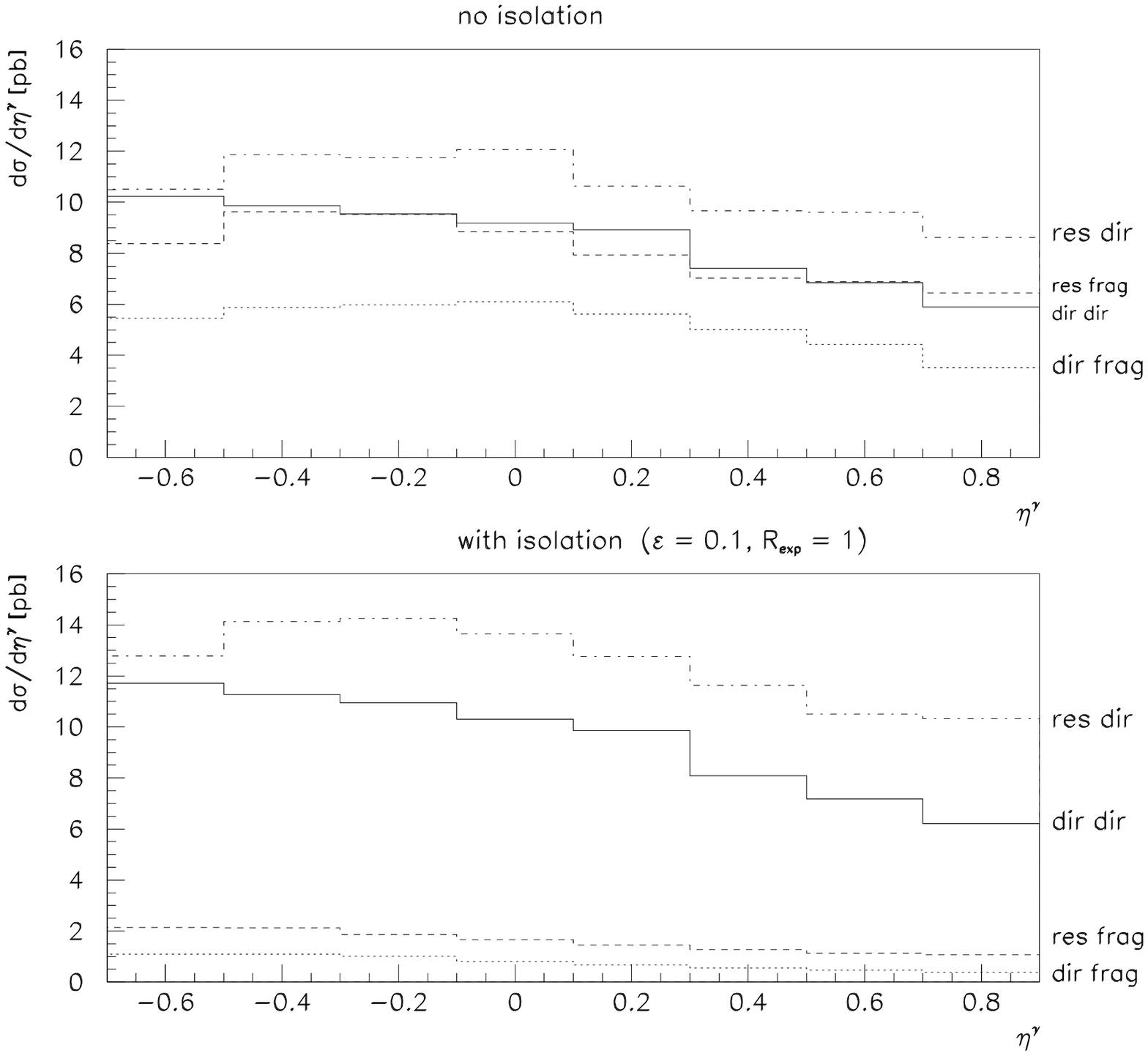,height=15cm}}
\end{center}
\caption{Relative magnitude of  
all four components contributing to $d\sigma^{ep\to\gamma X}/d\eta^{\gamma}$
for the scale choice $\mu=M=M_F=p_T^{\gamma}$.}
\label{isononiso}
\end{figure}

Fig.~\ref{isofull/frag} shows the rapidity distribution of the full 
cross section before and after isolation. 
As already mentioned in section \ref{iso}, we used the isolation cuts 
$E_{T\, max}=\epsilon \,p_T^{\gamma}$ with 
$\epsilon=0.1$ and $R_{\mathrm{exp}}$ = 1 to match those of the ZEUS collaboration. 
Fig.~\ref{isofull/frag} also shows the effect of isolation on the fragmentation 
part\footnote{By "fragmentation part" we mean the sum of the two subparts
"direct fragmentation" and "resolved fragmentation". Analogously, "resolved" 
denotes the sum of "resolved direct" and "resolved fragmentation".} separately. 
Isolation reduces 
the fragmentation component to about 6\% of the total isolated 
cross section.  \\
In Fig.~\ref{isononiso} the relative magnitude of  
all four components contributing to $d\sigma^{ep\to\gamma X}/d\eta^{\gamma}$
before and after isolation is shown. Note that isolation {\em increases}
the contributions with a direct photon in the final state slightly since there
the cut mainly acts on a negative term, which is the one 
where parton 5 is collinear to the photon. 
It should be emphasized that Figure~\ref{isononiso} has to be read with care 
since the individual
parts have no physical meaning and are very sensitive to scale changes. 
Nevertheless the dominance of the resolved direct part remains if we 
choose e.g. $\mu=M=M_F=p_T^{\gamma}/2$  or $2\,p_T^{\gamma}$.

\begin{figure}[htb]
\begin{center}
\mbox{\epsfig{file=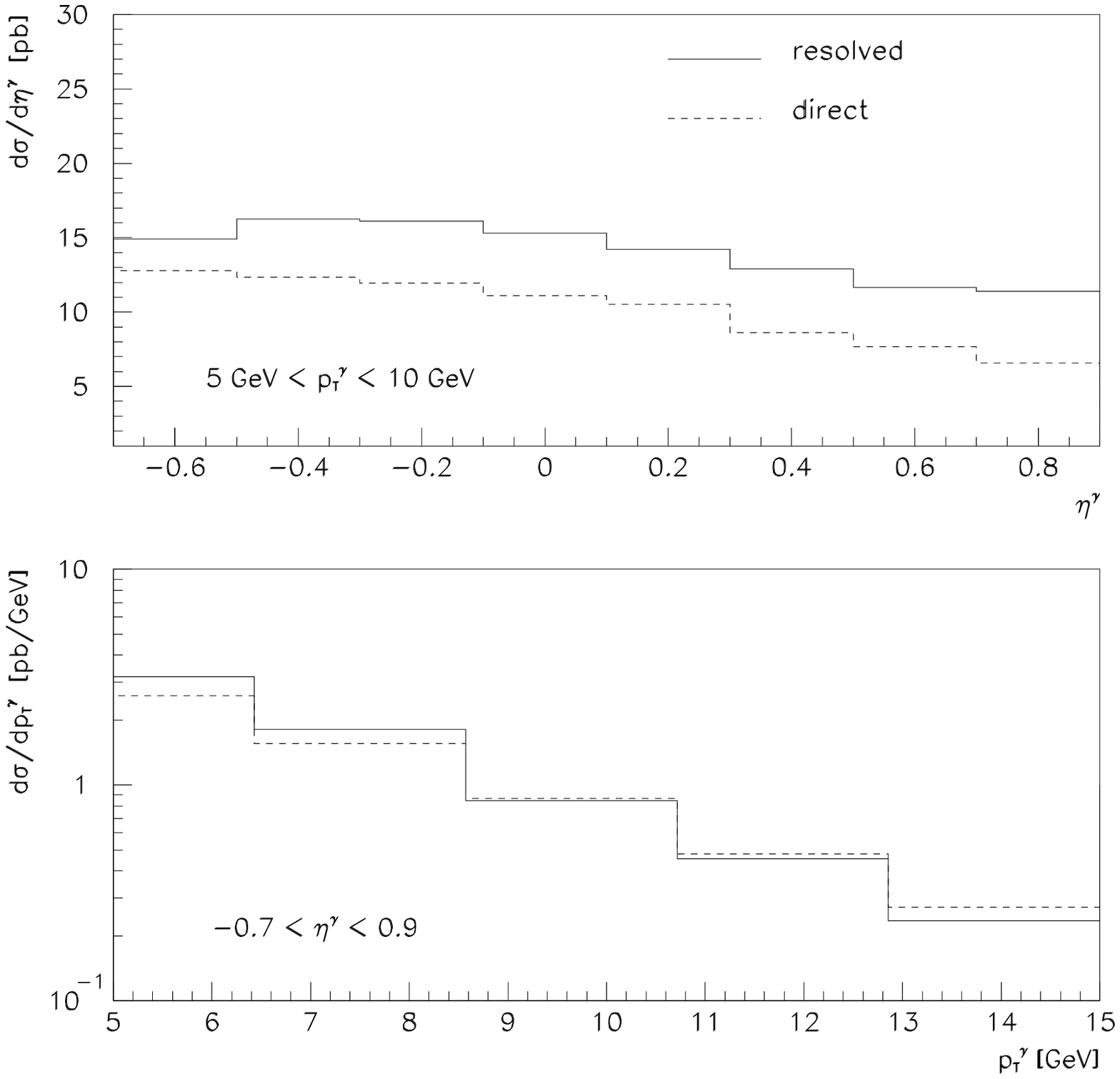,height=15cm}}
\end{center}
\caption{Comparison of contributions from resolved and direct photons
in the initial state for the photon rapidity and 
transverse momentum distribution, with isolation.} 
\label{resovsdir}
\end{figure}

Figure \ref{resovsdir} shows the relative magnitude 
of contributions from 
resolved and direct photons in the initial state to the isolated cross section. 
From the $p_T^{\gamma}$ distribution  one can conclude that 
the resolved part dominates 
the cross section for small values of $p_T^{\gamma}$ such that 
it would be useful to look at the photon rapidity distribution at 
$p_T^{\gamma}=5$\,GeV in order to discriminate between different parton 
distribution functions in the photon. 

\begin{figure}[htb]
\begin{center}
\mbox{\epsfig{file=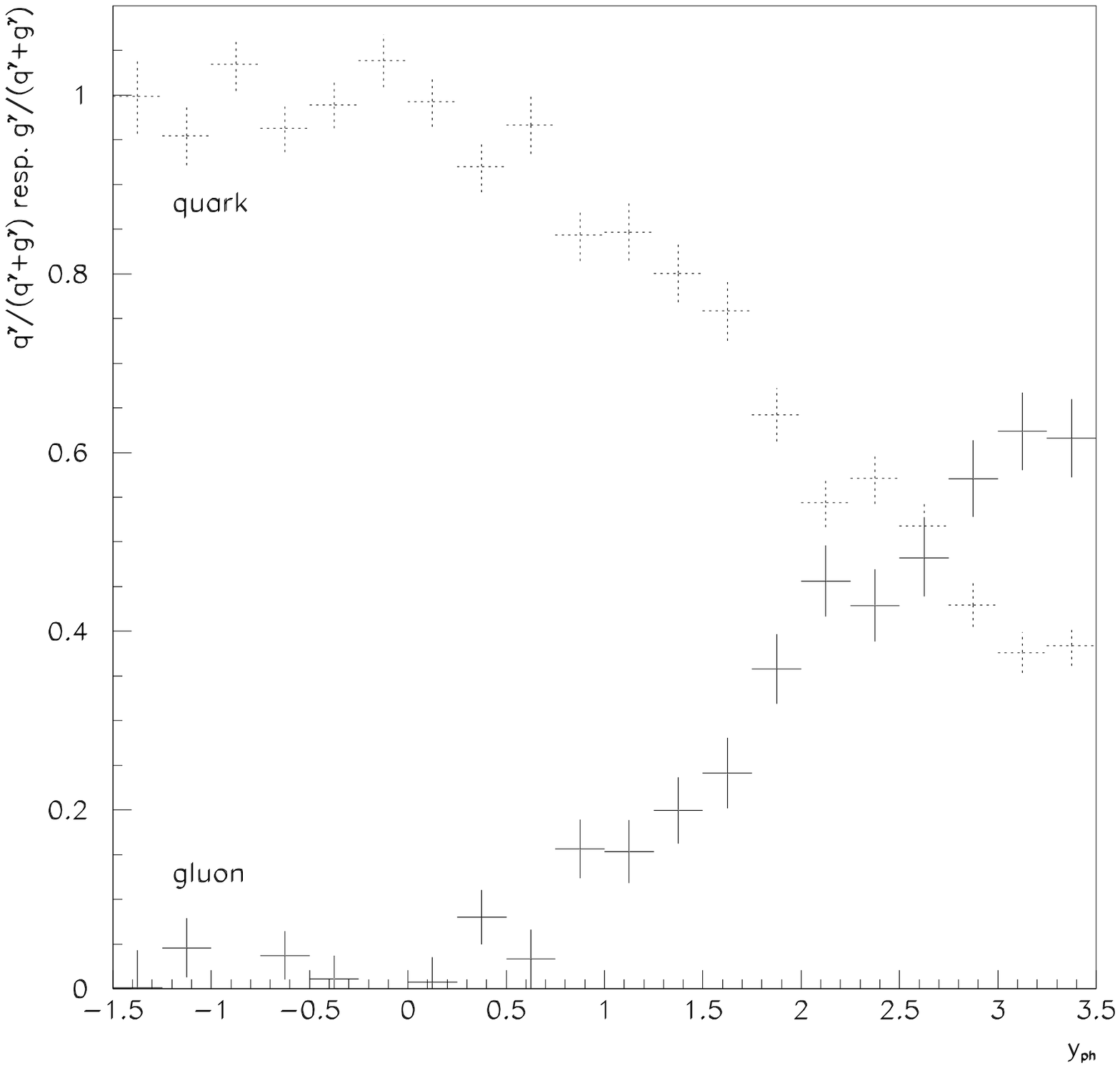,height=14cm}}
\end{center}
\caption{Ratio of the contribution from quark resp. gluon distributions in 
the photon to the full resolved part.}
\label{gluqua}
\end{figure}

Since the {\em gluon} distribution in the photon is of particular interest,
the sensitivity to the gluon in the photon is investigated in
Fig.~\ref{gluqua}. One can see that the  gluon 
distribution in the photon starts to become sizeable only for photon 
rapidities $\eta^{\gamma} > 1$ and dominates over the quark distribution 
for about $\eta^{\gamma} > 2.5$\,. 
Therefore the region of large photon rapidities and small 
photon $p_T$ is the one where the sensitivity to the gluon in the photon 
is largest. 
In order to test further the sensitivity to the gluon, 
we  increased the gluon distribution in the photon uniformly by 20\%. 
As can be anticipated from Fig.~\ref{gluqua}, the effect becomes 
sizeable only for $\eta^{\gamma} > 2$ and leads to an increase of the cross
section by about 10\% for  $\eta^{\gamma} > 2.5$\,.
We conclude that in the region $\eta^{\gamma} < 1$, 
there is basically no sensitivity to the gluon in the photon. 
 However, investigating the 
 photon + jet cross section instead of the inclusive case offers larger 
 possibilities to constrain the gluon in the photon since there one can vary 
 the photon {\em and} the jet rapidities in order to single out a kinematic 
 region where the sensitivity is large~\cite{jets}.

\begin{figure}[htb]
\begin{center}
\mbox{\epsfig{file=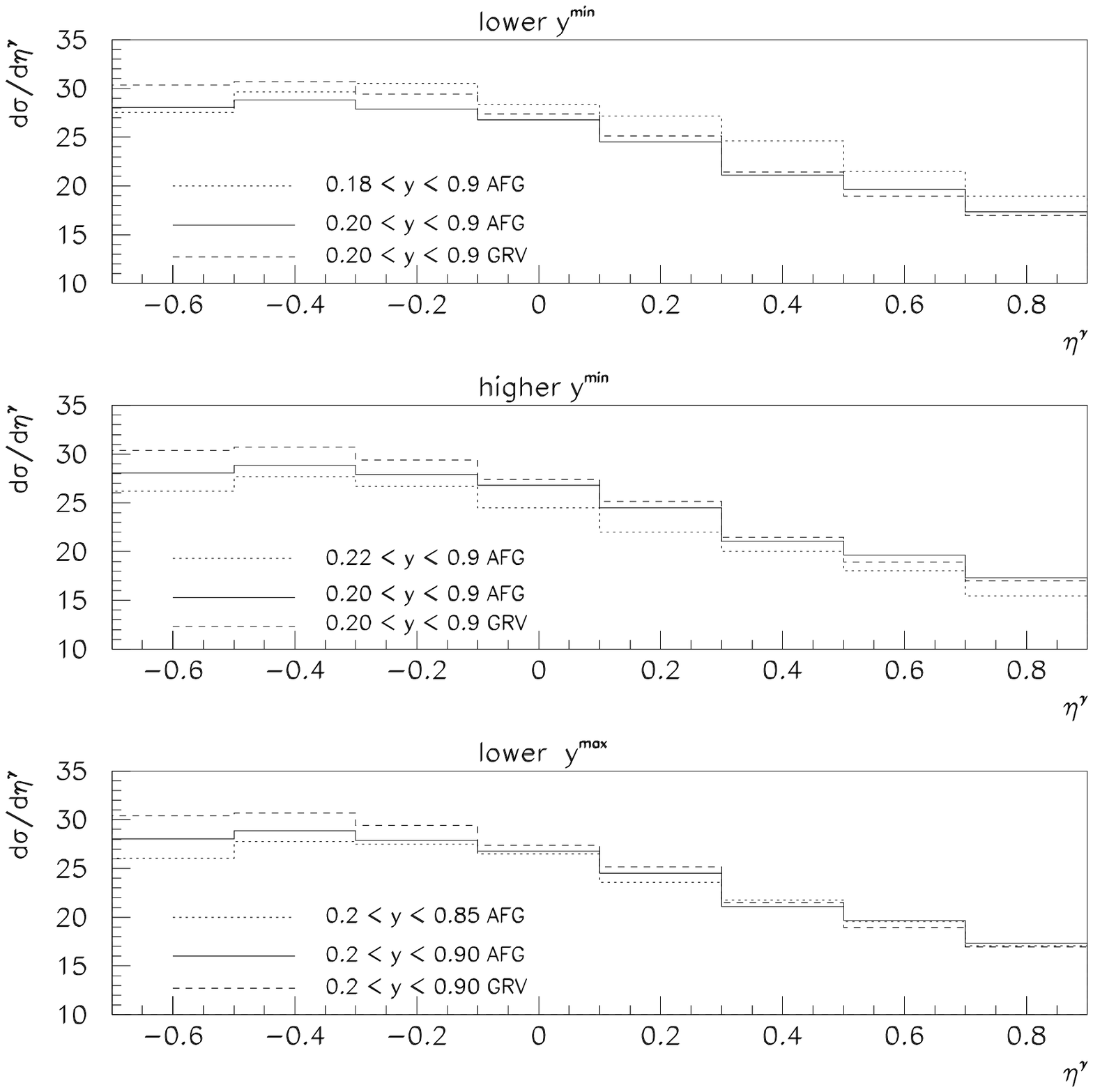,height=16.5cm}}
\end{center}
\caption{Photon rapidity distribution $d\sigma^{ep\to\gamma X}/d\eta^{\gamma}$ for 
isolated prompt photons integrated  
over 5 GeV $<p_T^{\gamma}<$ 10 GeV and different lower bounds on $y$. 
Solid line: $0.2<y<0.9$ with AFG photon structure functions, dotted line: 
bounds on $y$ changed by about 10\%, 
dashed line: $0.2<y<0.9$ with GRV photon structure functions}
\label{varyeta}
\end{figure}

Figure \ref{varyeta} shows  the effect of a ten percent uncertainty 
in the "true" bounds of the photon energy $y$.
One can see that a change of the lower bound on $y$ has a large effect, 
in particular at large photon rapidities. This comes from the fact that 
the Weizs\"acker-Williams distribution is large and steeply falling at 
small $y$. Increasing the lower bound on $y$ therefore removes a large
fraction of the direct events with lower energy initial photons. 
($y=x_e$ for the direct events and large $\eta^{\gamma}$ correspond to 
small $x_e$.) At large photon rapidities the spread  due to the use of
different parton distribution functions for the photon is smaller than the one 
caused by a 10\% variation of the lower bound on $y$. On the other hand, 
the region of large photon rapidities is of special interest since there the 
gluon in the photon is dominating. Therefore a small experimental error
in the reconstruction of the "true" photon energy is crucial in 
order to be able to discriminate between different sets of 
parton distribution functions in the photon.

It has been tested that the effect of using different {\em proton}
distribution functions -- for example the CTEQ4M or the MRST1
set of proton distribution functions  -- is of the order of 3\% at most. 
In all photon rapidity bins the spread is smaller than the one caused by different
sets of photon distribution functions (which is about 10\% at small photon 
rapidities, see e.g. Fig.~\ref{ygamma}). Thus a discrimination between different
sets of photon distribution functions should be possible with the forthcoming 
full 1996-2000 data set analysis, 
where the errors on the data are expected to be small enough. 


\clearpage

\subsection{Comparison with ZEUS data}

\begin{figure}[hb]
\begin{center}
\mbox{\epsfig{file=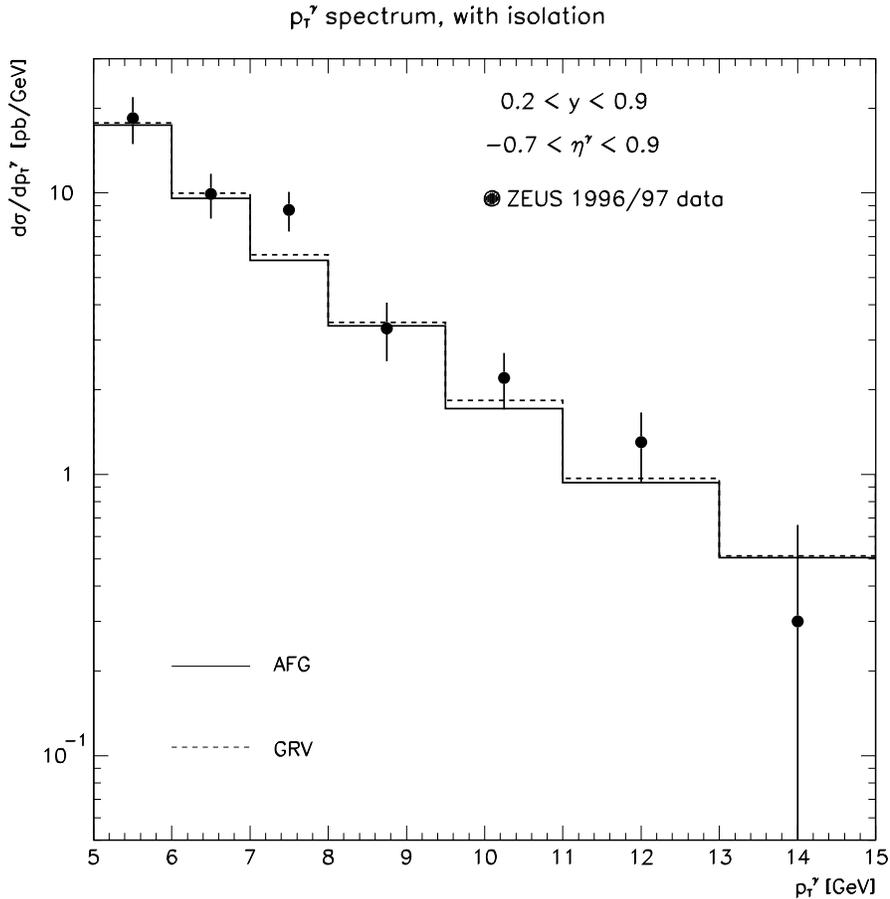,height=13cm}}
\end{center}
\caption{Comparison to ZEUS data of photon $p_T$ distribution 
$d\sigma^{ep\to\gamma X}/dp_T^{\gamma}$ for isolated prompt photons.
Results for two different sets of parton distributions in the photon are shown.}
\label{ptgamma}
\end{figure}

\begin{figure}[htb]
\begin{center}
\mbox{\epsfig{file=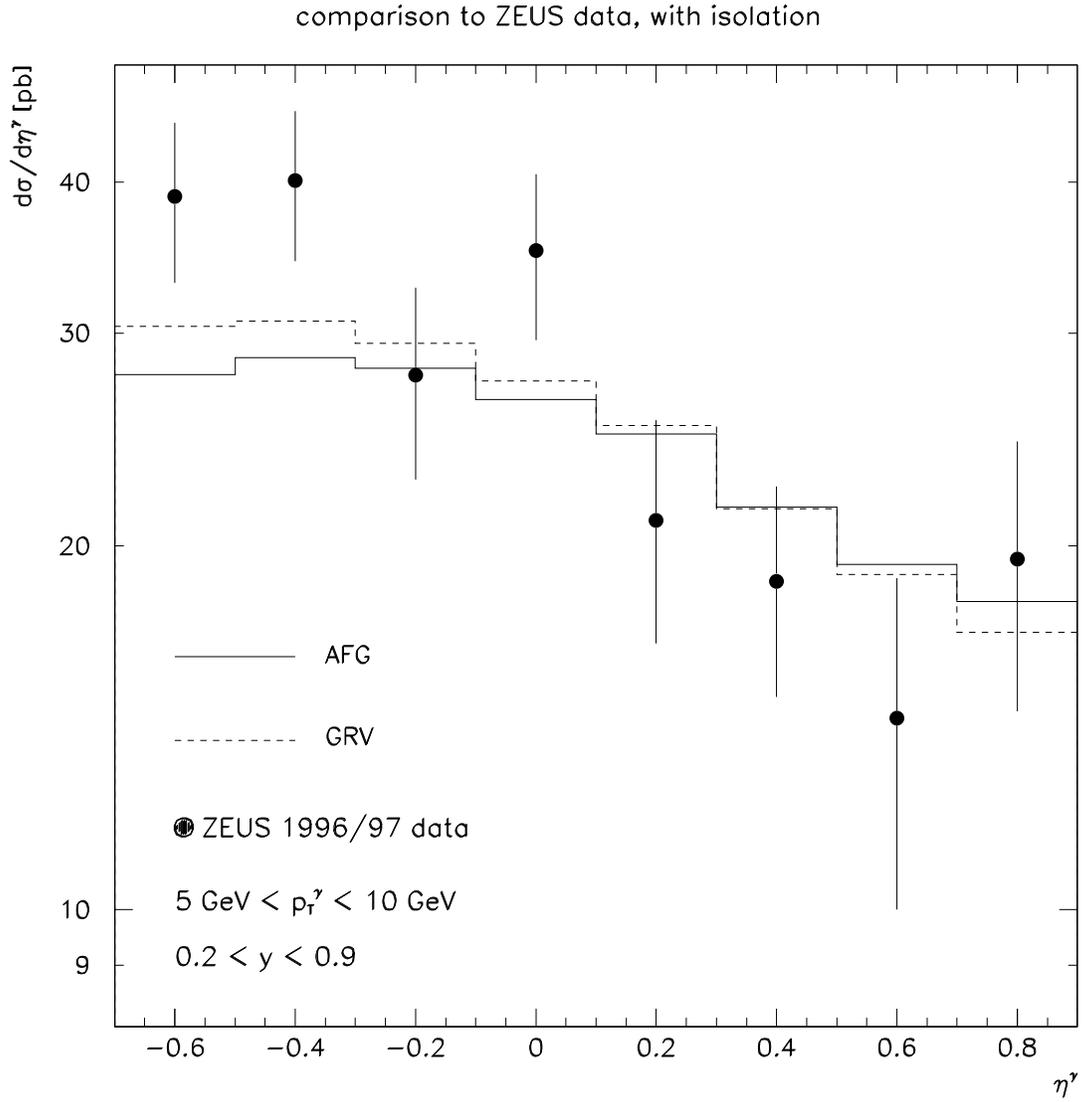,height=16cm}}
\end{center}
\caption{Comparison to ZEUS data of photon rapidity distribution 
$d\sigma^{ep\to\gamma X}/d\eta^{\gamma}$ for 
isolated prompt photons.}
\label{ygamma}
\end{figure}

In this section we compare our results to the ZEUS 1996-97 data
on inclusive prompt photon photoproduction~\cite{Breitweg:2000su}. 
Figures~\ref{ptgamma} and~\ref{ygamma} show the photon $p_T$ 
and rapidity distributions with 
AFG resp. GRV sets of structure functions for the photon. 
For the $p_T$ distribution the agreement between data and theory
is quite good. 
In the rapidity distribution (Fig.~\ref{ygamma}) the data fluctuate 
a lot, such that the agreement is still satisfactory. 
However, it seems that 
theory underpredicts the data in the backward region, whereas 
the theoretical prediction tends to be higher at large photon rapidities. 
The curves of Gordon~\cite{Gordon} and Krawczyk/Zembrzuski~\cite{Krawczyk}
given in~\cite{Breitweg:2000su}  also show this trend. 
At high $\eta^{\gamma}$ the reason for the difference could be that 
the isolation cut in the experiment removes more events 
than in the theoretical (parton level) simulation, as discussed in
section~\ref{iso}.  

\begin{figure}[htb]
\begin{center}
\mbox{\epsfig{file=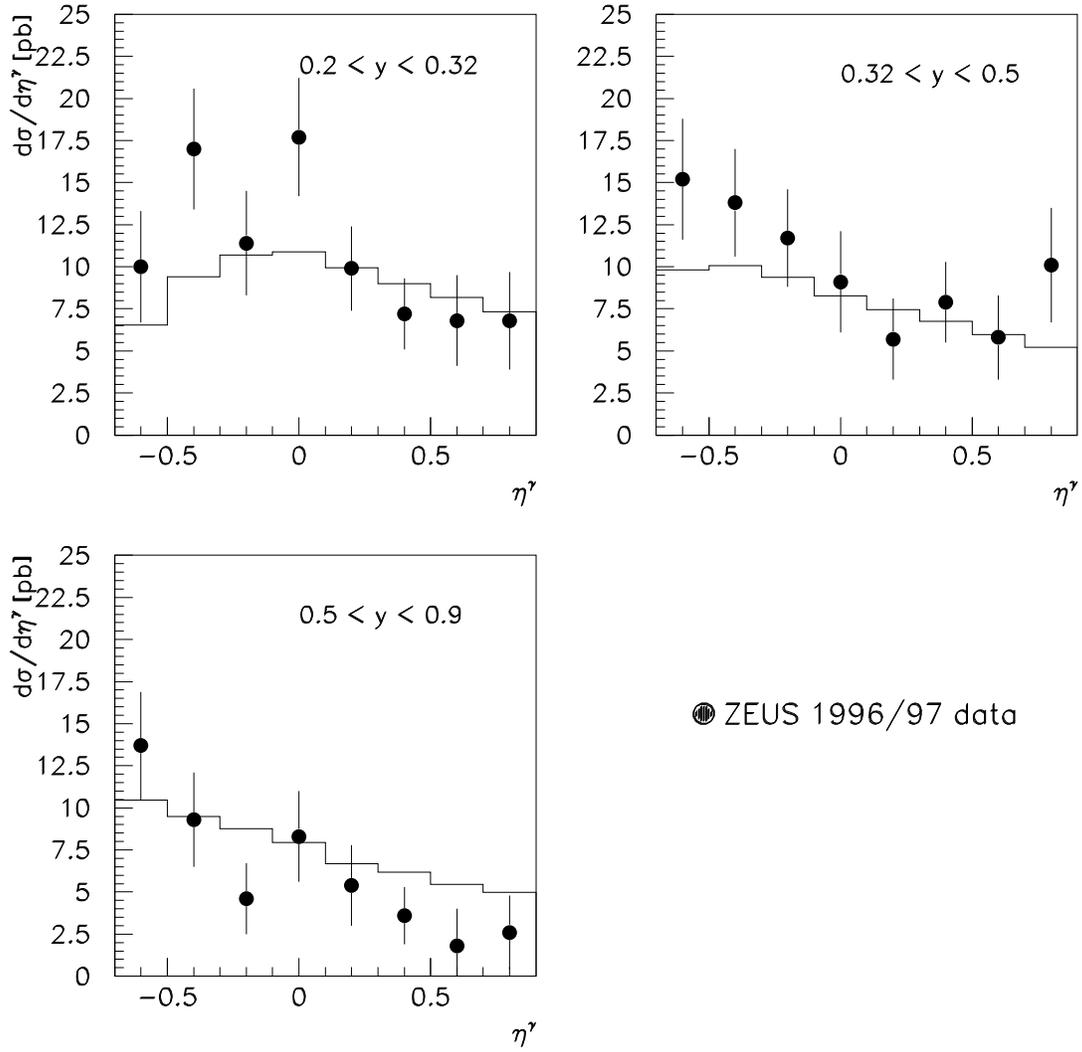,height=16cm}}
\end{center}
\caption{Photon rapidity distribution $d\sigma^{ep\to\gamma X}/d\eta^{\gamma}$  
integrated  over 5 GeV$<p_T^{\gamma}<$10 GeV and  different subdivisions 
of photon energies: (a) $0.2<y<0.32$, (b) $0.32<y<0.5$, (c) $0.5<y<0.9$.}
\label{sliceeta}
\end{figure}

Figure~\ref{sliceeta} shows that the discrepancy between theory and 
data at low $\eta^{\gamma}$ comes mainly from the domain of small 
photon energies, whereas the discrepancy at large $\eta^{\gamma}$ is only 
present in the range of large photon energies. Note that at large 
$\eta^{\gamma}$ and large $y$ the resolved part dominates and the underlying 
event could have a large multiplicity. Therefore the isolation 
criterion could also cut on the non-fragmentation 
contributions as discussed in section~\ref{iso}.   

\clearpage

\section{Conclusions}

We have presented a program for prompt photon photoproduction which includes 
the full next-to-leading order corrections to all contributing subparts. 
It is a general purpose code of partonic event generator type and as such very
flexible. 

We used it to study the possibility to constrain the quark and gluon 
distributions in the photon. It turned out that the sensitivity to 
the {\em gluon} distribution in the photon is negligible in the 
rapidity range $-0.7<\eta^{\gamma}<0.9$ studied by ZEUS. 

A discrimination between the AFG/GRV sets of parton distributions in the photon 
is not possible with the present experimental errors on the ZEUS 1996/97 data.
However, a forthcoming analysis of all 1996-2000 data announced by the ZEUS
collaboration will drastically improve this situation. 

We have shown that the cross section is very sensitive to small variations of
the photon energy range. Therefore a good control of the experimental error on
the photon energy fraction $y$ (reconstructed experimentally from the 
Jacquet-Blondel variable $y_{JB}$) will be crucial for future comparisons. 

Despite the large fluctuations of the data, one can say that there is 
a trend that theory overpredicts the data in the forward region. 
The reason might be that the isolation cut imposed at partonic level
in the perturbative QCD calculation does not have the same effect as 
the experimental one. 
If the experimental cut is too stringent, a large fraction of the 
hadronic energy in the isolation cone may come from underlying events,
such that experimentally a larger number of events is rejected. 
We gave a rough estimate of the  underlying events to be
expected in the isolation cone. 

The possibilities offered by the study of photon + jet photoproduction 
will be investigated in a forthcoming publication~\cite{jets}. 

\vspace*{8mm}       

{\bf\Large Acknowledgements}

\vspace*{5mm}

We would like to thank P.~Bussey from the ZEUS collaboration for 
helpful discussions. 
G.H. would like to thank the LAPTH for its continuous  hospitality. 
This work was supported by the EU Fourth Training Programme  
''Training and Mobility of Researchers'', network ''Quantum Chromodynamics
and the Deep Structure of Elementary Particles'',
contract FMRX--CT98--0194 (DG 12 - MIHT).


\end{document}